\documentclass[12pt]{article}
\usepackage{cite,amssymb,amsmath,authblk}
\usepackage{bbold}
\usepackage[dvips]{color}
\begin{document}
\title{\bf Fermions in the 6D Standing Wave Braneworld with Real Scalar Field}
\author{\bf Pavle Midodashvili \thanks{pmidodashvili@yahoo.com}}
\affil{Ilia State University, 3/5 Cholokashvili Ave., Tbilisi 0162, Georgia}
\date{}
\maketitle
\begin{abstract}
In the article, we investigate localization problem for spinor fields within the 6D standing wave braneworld with the bulk real scalar field, introduced earlier in \cite{PM_2020}, and explicitly show that there is no normalizable fermion field zero mode trapped on the brane.
\vskip 0.3cm
Keywords: brane; standing waves; trapping of particles; zero modes

PACS numbers: 04.50.-h; 04.20.Jb; 11.25.-w
\end{abstract}

\vskip 0.5cm

Braneworld models \cite{AHDD,AAHDD,G1,G2,RS1,RS2} are very useful in addressing several unsolved problems in high energy physics (e.g. the hierarchy problem, the nature of flavor, etc.)(for reviews see \cite{Rub,Lan,Man,Maa-Koy}). Along with many static braneworld models, there are also dynamic braneworld models, appearing mainly in cosmological studies, which involve time-dependent metrics and source matter fields, as well as branes with tensions varying in time \cite{Gut-Str,Kru-Mye-Pee,Iva-Sin,Bur-Que-Rab-Tas-Zav}.

Localization problem is the key issue in realization of a braneworld scenario. For reasons of economy and stability it would be nice to have a universal gravitational trapping mechanism for all fields. In the 5D and 6D braneworld models proposed in \cite{Gog-Sin1,Gog-Her-Mal,Gog-Her-Mal-Mor-Nuc,Gog-Her-Mal-Mor,Sou-Sil-Alm,PM},
the brane was generated by standing waves of gravitational and phantom-like bulk scalar field, and the universal gravitational trapping of zero modes of all kinds of matter fields on the brane was provided by rapid oscillations of the waves   \cite{PM,GMM1,GMM2,GMM3,G3,Gog-Sakh-Tukh}. Later in \cite{GM_RealScField1} it was introduced another dynamic 5D anisotropic braneworld generated by standing waves of the gravitational and real scalar fields, instead of the phantom-like scalar fields of \cite{Gog-Sin1,Gog-Her-Mal,Gog-Her-Mal-Mor-Nuc,Gog-Her-Mal-Mor,Sou-Sil-Alm,PM}. The metric of that model had horizons in the bulk, and in the case of small amplitudes of standing waves matter was actually trapped on the brane of the width equal to the horizon size, while in the case of large amplitudes of standing waves the zero modes of all kinds of matter fields were trapped on the brane of the width much less than the horizon size \cite{GM_RealScField2,GM_RealScField3}.

Recently we have introduced the new non-static 6D braneworld model generated by standing gravitational waves coupled to bulk real scalar field \cite{PM_2020}, which differs in underlying metric from that considered in \cite{SCA}, where authors also studied the 6D standing-wave braneworld with normal matter as source.

In \cite{PM_2020}, the 6D standing wave braneworld with real scalar field is introduced with the metric \textit{ansatz}
\begin{equation}\label{Metric}
  d{s^2} = \frac{{{e^S}}}{{{{f}^{{\frac{3}{4}}}}}}\left( {d{t^2} - d{r^2}} \right)\\ - {f^{\frac{1}{2}}}\left[ {{e^u}\left( {d{x^2} + d{y^2} + d{z^2}} \right) + {e^{ - 3u}}d{\theta ^2}} \right],
\end{equation}
where the metric function $f$ is defined as
\begin{equation}\label{f}
  f=1+a\left|r\right|,
\end{equation}
with parameter $a$ being a positive constant.
The determinant of the metric (\ref{Metric}) is
\begin{equation}\label{DeterminantOfMetric}
  \sqrt {-g}  = {f^{\frac{1}{4}}}{e^S}.
\end{equation}
The metric functions $u$ and $S$ are solutions to 6D Einstein equations and have the following form \cite{PM_2020}:
\begin{eqnarray}
\label{u_s} \nonumber
  &&u\left( {t,r} \right) = A\sin \left( {\omega t} \right){J_0}\left( {\frac{\omega }{a}f} \right), \\
 &&S\left(r\right)=3A^2\left(\frac{\omega}{a}f\right)
 ^2\left[J_0^2\left(\frac{\omega}{a}f\right)+
 J_1^2\left(\frac{\omega}{a}f\right)
 -\frac{J_0\left(\frac{\omega}{a}f\right)J_1\left(
 \frac{\omega}{a}f\right)}{\frac{\omega}{a}f}\right]-B,
\end{eqnarray}
where $A$ and $\omega$ are some constants, $J_0$ and $J_1$ are Bessel functions of the first kind, and the integration constant $B$ is fixed as
\begin{equation}\label{Constant_B}
   B= 3A^2\frac{\omega^2 }{a^2}{{{J_1^2}\left( {\frac{\omega }{a}} \right)}}.
\end{equation}

There is also the bulk real scalar field standing wave, the solution to the 6D Klein-Gordon equations in the background metric (\ref{Metric}) \cite{PM_2020},
\begin{equation}\label{phi}
  \phi \left( {t,r} \right) = \sqrt 3 {M^2}A\cos \left( {\omega t} \right){J_0}\left( {\frac{\omega }{a}f} \right),
\end{equation}
where $M$ denotes the 6D fundamental scale, which is related to the 6D Newton's constant $G$ by the relation $G=\frac{1}{8\pi M^{4}}$. The scalar field $\phi \left( {t,r} \right)$ and the metric function  $u\left( {t,r} \right)$ are oscillating $\pi /2$ out of phase in time, but have similar dependence on the spatial coordinate $r$.

Bulk standing gravitational wave $u\left( {t,r} \right)$, given by (\ref{u_s}), has the frequency $\omega$ and the amplitude $A {J_0}\left( {\frac{\omega }{a}f} \right)$. In the extra dimension $r$ the wave is bounded by the brane at $r=0$, and it decays at infinities $r \to \pm \infty$. The standing wave amplitude vanishes at the points where the Bessel function ${J_0}\left( {\frac{\omega }{a}f} \right)$ is zero, these points correspond to nodes of the bulk standing wave. In the model it is also imposed the quantization condition
\begin{equation}\label{QuntizationCondition}
  \frac{\omega}{a}=Z_n,
\end{equation}
where the constant $Z_n$ denotes the $n$-th zero of the Bessel function $J_0$, and $n \ge 1$ is some fixed integer. This condition guarantees that the 5D-brane is located at the origin $r=0$ of the infinite spacelike extra dimension $r$ orthogonal to the brane \cite{PM_2020}. Among four spatial coordinates of the 5D-brane, three of them, $x$, $y$ and $z$, denote the ordinary dimensions of our universe, while the remaining one, $\theta$, is assumed to be compact, curled up to the unobservable sizes for the present energies (hybrid compactification), and it is normalized as
\begin{equation}\label{ThetaNormalization}
  0\le\frac{\theta}{\theta_0}\le 2\pi,
\end{equation}
where $\theta_0$ is some fixed length scale.

Using (\ref{QuntizationCondition}) the parameter (\ref{Constant_B}) can be written as
\begin{equation}\label{Constant_B_InTermsOf_Z}
  B= 3A^2{Z_n^2}{{{J_1^2}\left( {Z_n} \right)}}.
\end{equation}

It was shown in \cite{PM_2020} that in the case of large parameter $B$,
\begin{equation}\label{LargeParameter_B}
  B \gg 1,
\end{equation}
the metric functions $S$ and $u$ defined by (\ref{u_s}) obey the relation:
\begin{equation}\label{Relation_u_S}
  \left |\frac{u \left(t,r\right)}{S\left(r\right)} \right | \ll 1,
\end{equation}
which holds for any $t$ and $r$.
Then the width of the brane along the $r$ extra dimension,
\begin{equation}\label{BraneWidth_Delta}
  \Delta \sim \frac{1}{aB},
\end{equation}
which is located at the origin of the $r$ extra dimension, is determined by the metric function $S\left(r\right)$ \cite{PM_2020}, and matter fields are dynamically trapped on the brane by the bulk standing wave of gravity.

Close to the brane at $r=0$ the metric (\ref{Metric}) has the following approximate form:
\begin{eqnarray} \label{Metric_Approximation_r=0}
\nonumber
  g_{tt} &=& -g_{rr} \approx 1+\frac{3\left|r\right|}{\Delta},\\
  g_{xx}&=&g_{yy}=g_{zz}\approx -1+O\left(\frac{\left|r\right|}{\Delta B}\right), \\
  \nonumber
  g_{\theta \theta}&\approx& -1+O\left(\frac{\left|r\right|}{\Delta B}\right).
\end{eqnarray}
Then the zero mode wavefunction $\Phi\left(x^A\right)$ of a free brane particle with the energy-momentum
\begin{equation}\label{}
  P_\mu \ll \frac{1}{\Delta},
\end{equation}
can be factorized as:
\begin{equation}\label{}
  \left.\Phi\left(x^A\right)\right|_{r \to 0}\approx e^{-iP_\mu x^\mu}\varphi\left({r,\theta} \right),
\end{equation}
where Greek indices numerate 4D coordinates $t$, $x$, $y$ and $z$, and the scalar function $\varphi\left({r,\theta} \right)$ is the extra dimension factor of the wavefunction near the brane. It is obvious from (\ref{Metric_Approximation_r=0}) that, in the case (\ref{LargeParameter_B}), the Lorentz violation terms in $g_{xx}$, $g_{yy}$, $g_{zz}$ and $g_{\theta \theta}$ are much less than KK corrections, $\approx 1/\Delta$, in $g_{tt}$ and $g_{rr}$.

In the article \cite{PM_2020}, it was shown that in the case of large parameter $B$, (\ref{LargeParameter_B}), there are two different physical limits realizing  the pure gravitational trapping of classical particles and light and also of massless scalar fields on the brane. In the first physical limit the standing wave amplitudes have to be sufficiently large, while in the second one the amplitudes of standing waves can be small enough. And it was explicitly shown in \cite{PM_2020} that in both cases particles and light, as well as massless scalar fields are dynamically trapped on the brane by the pressure of bulk standing waves of gravity and real scalar field.

Now let us consider the problem of localization of 6D spin $\frac{1}{2}$ fermion field in this braneworld model. The starting action is the 6D Dirac action of free massless spin $\frac{1}{2}$ fermion
\begin{equation}\label{DiracAction}
  S_{\frac{1}{2}}=\int d^6x\sqrt{-g}i\overline{\Psi}\left(x^A\right)
  \Gamma^MD_M\Psi\left(x^A\right).
\end{equation}

The curved gamma matrices , $\Gamma^A$, which obey the relations
\begin{equation}\label{6D_GammaMatrRel}
  \{ \Gamma ^ A , \Gamma ^ B \}=2g^{AB},~~(A,B,...=t,x,y,z,r,\theta),
\end{equation}
can be chosen as:
\begin{equation}\label{6D_GammaMatrRelToFlatMatr}
  \Gamma^A=h_{\bar{A}}^A\Gamma^{\bar{A}},~~~
  \Gamma^{\bar{A}}=\left(\gamma^t,\gamma^x,
  \gamma^y,\gamma^z,\gamma^r,\gamma^{\theta}\right),
\end{equation}
where the indices $\bar{A}$, $\bar{B}$, ... refer to 6D local Lorentz (tangent) frame and $A$, $B$, ... stand for 6D spacetime coordinates $t$, $x$, $y$, $z$, $r$, $\theta$.

The \textit{vielbein} for the metric (\ref{Metric}),
\begin{eqnarray}\label{Vielbein}
  &&h_A^{\bar{A}}=diag\left(
  \frac{e^{\frac{S}{2}}}{f^{\frac{3}{8}}},
  f^{\frac{1}{4}}e^{\frac{u}{2}},
  f^{\frac{1}{4}}e^{\frac{u}{2}},
  f^{\frac{1}{4}}e^{\frac{u}{2}},
  \frac{e^{\frac{S}{2}}}{f^{\frac{3}{8}}},
  f^{\frac{1}{4}}e^{-\frac{3u}{2}}
  \right),\\
  \nonumber
  &&h^{\bar{A}A}=g^{AB}h_B^{\bar{A}},
  ~~h_{\bar{A}}^A=\eta_{\bar{A}\bar{B}}h^{\bar{B}A},
  ~~h_{\bar{A}A}=\eta_{\bar{A}\bar{B}}h_A^{\bar{B}},
\end{eqnarray}
is introduced through the conventional definition:
\begin{equation}\label{VielbeinDefinition}
  g_{AB}=\eta_{\bar{A}\bar{B}}h_A^{\bar{A}}h_B^{\bar{B}}.
\end{equation}

According to (\ref{6D_GammaMatrRelToFlatMatr}), the curved spacetime gamma matrices are related to Minkowskian ones by the following expressions:
\begin{eqnarray}
  \nonumber
  \Gamma^t&=&f^{\frac{3}{8}}e^{-\frac{S}{2}}\gamma^t, \\
  \nonumber
  \Gamma^x&=&f^{-\frac{1}{4}}e^{-\frac{u}{2}}\gamma^x, \\
  \Gamma^y&=&f^{-\frac{1}{4}}e^{-\frac{u}{2}}\gamma^y,\\
  \nonumber
  \Gamma^z&=&f^{-\frac{1}{4}}e^{-\frac{u}{2}}\gamma^z, \\
  \nonumber
  \Gamma^r&=&f^{\frac{3}{8}}e^{-\frac{S}{2}}\gamma^r, \\
  \nonumber
  \Gamma^{\theta}&=&f^{-\frac{1}{4}}e^{\frac{3u}{2}}
  \gamma^{\theta}.
\end{eqnarray}

In the 6D Dirac action (\ref{DiracAction}) the covariant derivatives are:
\begin{equation}\label{CovariantDerivatives}
  D_A=\partial_A+\frac{1}{4}\Omega_A^{\bar{B}\bar{C}}
  \Gamma_{\bar{B}}\Gamma_{\bar{C}},
\end{equation}
with $\Omega_M^{\bar{M}\bar{N}}$ denoting the spin-connection,
\begin{eqnarray}
\nonumber
  \Omega_M^{\bar{M}\bar{N}}=-\Omega_M^{\bar{N}\bar{M}}&=&
  \frac{1}{2}\left[h^{N\bar{M}}\left(\partial_Mh_N^{\bar{N}}-
  \partial_Nh_M^{\bar{N}}\right)
  -h^{N\bar{N}}\left(\partial_Mh_N^{\bar{M}}-
  \partial_Nh_M^{\bar{M}}\right)\right. \\
  &&~~~~~~~~\left.-h_M^{\bar{A}}h^{P\bar{M}}h^{Q\bar{N}}
  \left(\partial_Ph_{Q\bar{A}}-\partial_Qh_{P\bar{A}}\right)\right].
\end{eqnarray}
The non-vanishing components of the spin-connection in the background (\ref{Metric}) are:
\begin{eqnarray} \label{Spin-connection}
\nonumber
\Omega_t^{\bar{t}\bar{r}} &=& -\frac{S^{\prime}}{2}+\frac{3a\,\rm{sgn}\left(r\right)}{8f},\\
\nonumber
\Omega_x^{\bar{x}\bar{r}} &=& \Omega_y^{\bar{y}\bar{r}}=\Omega_z^{\bar{z}\bar{r}}= -\left(\frac{u^{\prime}}{2}f^{\frac{5}{8}}+\frac{a\,\rm{sgn}
\left(r\right)}{4f^{\frac{3}{8}}}\right)e^{-\frac{S}{2}+\frac{u}{2}},
\\
  \Omega_x^{\bar{x}\bar{t}} &=& \Omega_y^{\bar{y}\bar{t}}=\Omega_z^{\bar{z}\bar{t}}= \frac{\dot{u}}{2}f^{\frac{5}{8}}e^{-\frac{S}{2}+\frac{u}{2}}, \\
  \nonumber
  \Omega_{\theta}^{\bar{\theta}\bar{t}}&=& -\frac{3\dot{u}}{2}f^{\frac{5}{8}}e^{-\frac{S}{2}-\frac{3u}{2}}, \\
  \nonumber
  \Omega_{\theta}^{\bar{\theta}\bar{r}} &=& \left(\frac{3u^{\prime}}{2}f^{\frac{5}{8}}-\frac{a\,\rm{sgn}
  \left(r\right)}{4f^{\frac{3}{8}}}\right)e^{-\frac{S}{2}-\frac{3u}{2}},
\end{eqnarray}
where dots and primes denote derivatives with respect to the time $t$ and the extra coordinate $r$ respectively.

Using (\ref{Spin-connection}), the covariant derivatives (\ref{CovariantDerivatives}) can be written as:
\begin{eqnarray}
  D_t &=& \partial_t+\frac{1}{2}\left[\frac{S^{\prime}}{2}-\frac{3a\,\rm{sgn}\left(r\right)}{8f}\right]
  \gamma_r\gamma_t, \\
  D_x&=&\partial_x+f^{\frac{5}{8}}e^{-\frac{S}{2}+\frac{u}{2}}
  \left[\left(\frac{u^{\prime}}{4}+\frac{a\,\rm{sgn}\left(r\right)}{8f}\right)\gamma_r-\frac{\dot{u}}{4}
  \gamma_t\right]\gamma_x, \\
  D_y&=&\partial_y+f^{\frac{5}{8}}e^{-\frac{S}{2}+\frac{u}{2}}
  \left[\left(\frac{u^{\prime}}{4}+\frac{a\,\rm{sgn}\left(r\right)}{8f}\right)\gamma_r-\frac{\dot{u}}{4}
  \gamma_t\right]\gamma_y, \\
  D_z&=&\partial_z+f^{\frac{5}{8}}e^{-\frac{S}{2}+\frac{u}{2}}
  \left[\left(\frac{u^{\prime}}{4}+\frac{a\,\rm{sgn}\left(r\right)}{8f}\right)\gamma_r-\frac{\dot{u}}{4}
  \gamma_t\right]\gamma_z, \\
  D_r &=& \partial_r, \\
  D_{\theta}&=&\partial_{\theta}+f^{\frac{5}{8}}e^{-\frac{S}{2}-\frac{3u}{2}}
  \left[\left(-\frac{3u^{\prime}}{4}+\frac{a\,\rm{sgn}\left(r\right)}{8f}\right)\gamma_r+\frac{3\dot{u}}{4}
  \gamma_t\right]\Gamma_{\theta}.
\end{eqnarray}
The 6D Dirac equation ,
\begin{equation}\label{DiracEquation}
  i\gamma^AD_A\Psi=0,
\end{equation}
reduces to
\begin{equation}\label{DiracEquationReduced}
  \left[\gamma^{{t}}\partial_t+\frac{e^{\frac{S}{2}-
  \frac{u}{2}}}{f^{\frac{5}{8}}}\left(\gamma^{{i}}\partial_i+
  e^{2u}\gamma^{\theta}\partial_{\theta}\right)+
  \gamma^{{r}}\left(\partial_r+\frac{S^{\prime}}{4}+
  \frac{5a\,\rm{sgn}\left(r\right)}{16f}\right)\right]\Psi=0,
\end{equation}
where $\gamma^{{i}}\partial_i$ denotes the sum over $i=x, y, z$:
\begin{equation}\label{Sum}
  \gamma^{{i}}\partial_i=\gamma^{{x}}\partial_x+\gamma^{{y}}\partial_y+\gamma^{{z}}\partial_z~.
\end{equation}
Using (\ref{Relation_u_S}), the function $u$ can be ignored in comparison with the function $S$ and the equation gets the form:
\begin{equation}\label{ReducedDiracEqn_Approximation}
  \left[\gamma^{{t}}\partial_t+\frac{e^{\frac{S}{2}}}{f^{\frac{5}{8}}}\left(\gamma^{{i}}\partial_i+
  \gamma^{\theta}\partial_{\theta}\right)+
  \gamma^{{r}}\left(\partial_r+\frac{S^{\prime}}{4}+
  \frac{5a\,\rm{sgn}\left(r\right)}{16f}\right)\right]\Psi=0.
\end{equation}
In general the solution to this equation has non-trivial dependence on 4D, $r$ and $\theta$ coordinates. We look for the solution to this equation in the form:
\begin{equation}\label{Fermion_Ansatz}
  \Psi \left( {t,x,y,z,r,\theta } \right) = \psi \left( {{x^\mu }} \right)\sum\limits_{l,m} {{\alpha _m}\left(r \right){e^{il\frac{\theta }{{{\theta _0}}}}}},
\end{equation}
where 4D factor $\psi \left( {{x^\mu }} \right)$ of the spinor field wavefunction obeys the equation:
\begin{equation}\label{Spinor_4D-Factor_Equation}
  \gamma^{{\mu}}\partial_{\mu}\psi \left( {{x^\mu }} \right)=0.
\end{equation}
Let us use this decomposition of the 6D spinor wavefunction for two limiting regions: close to the brane, $r \to 0$, where the zeroth mode must be located, and far from the brane, $r \to \pm\infty$.

Close to the brane, $r \to 0$, we assume that 4D part of (\ref{Fermion_Ansatz}), $\psi\left(x^{\mu}\right)$, corresponds to the 4D zero mode Dirac spinor, and
\begin{equation}\label{Spinor_4D-Factor_Equation_OnBrane}
  \gamma^{{\mu}}\partial_{\mu}\psi \left( {{x^\mu }} \right)=0.
\end{equation}
In this region the equation (\ref{ReducedDiracEqn_Approximation}) reduces to the following asymptotic form
\begin{equation}\label{Equation_alpha_at_0}
  \left(\partial_r+\frac{S^{\prime}}{4}+
  \frac{5a\,\rm{sgn}\left(r\right)}{16f}\right)
  \alpha_0\left(r\right)=0~,
\end{equation}
which has the solution:
\begin{equation}\label{Solution_alpha_at_0}
  \alpha_0\left(r\right)=\frac{Ce^{-\frac{S}{4}}}{f^{\frac{5}{16}}}~,
\end{equation}
where $C$ is the integration constant.

Far from the brane, $r \to \pm\infty$, using the factorization (\ref{Fermion_Ansatz}), and assuming that at the infinities $r \to \pm\infty$
\begin{equation}\label{psi_at_infinity}
  \psi\left(x^{\mu}\right)=const~,
\end{equation}
for the extra dimension factor $\alpha_0\left(r\right)$ we get the same asymptotic equation (\ref{Equation_alpha_at_0}) with the same solution (\ref{Solution_alpha_at_0}).

To have a localized field on a brane, its 4D effective Lagrangian, which appears upon integration of corresponding 6D action over extra coordinates, must be non-vanishing and finite. But in our case this condition is not true for the zero mode spinor field action (\ref{DiracAction}). Indeed, putting the found solution $\alpha_0\left(r\right)$ (\ref{Solution_alpha_at_0}) into the 6D Dirac action (\ref{DiracAction}) and performing integration over extra coordinate $\theta$, the corresponding 4D effective action gets the following form:
\begin{equation}\label{Effective_ZeroMode_4D_Dirac_Action}
  S_{\frac{1}{2}}^{\left(0\right)}=
  2\pi\theta_0C^2\int d^4xi\overline{\psi}\left(x^\mu\right)\left[I_1\gamma^t\partial_t
  +I_2\gamma^i\partial_i\right]
  \psi\left(x^\mu\right),
\end{equation}
where
\begin{equation}\label{I1_andI2}
  I_1=\int\limits_{ - \infty }^\infty  {{f^{\frac{3}{8}}}{e^{ - \frac{S}{2}}}dr},~~~~~~~I_2=\int\limits_{ - \infty }^\infty  {{f^{\frac{1}{4}}}dr}.
\end{equation}
Taking into account (\ref{f}) and the asymptotics of the metric function $S\left(r\right)$ \cite{PM_2020},
\begin{equation}
\label{Asymptotics_Of_S_At_Infinity}
{\left. S \right|_{\left| r \right| \to \infty }} = 3Ba\left| r \right| + O\left(\frac{1}{\left| r \right|} \right)~,
\end{equation}
in (\ref{Effective_ZeroMode_4D_Dirac_Action}), the first integral, $I_1=\int\limits_{ - \infty }^\infty  {{f^{\frac{3}{8}}}{e^{ - \frac{S}{2}}}dr}$, is convergent (it gives some concrete finite real number), but the second integral, $I_2=\int\limits_{ - \infty }^\infty  {{f^{\frac{1}{4}}}dr}$, is divergent, which means that the zero mode is not localized on the brane.

To conclude, in this paper, we have investigated localization problem for fermion fields within the 6D standing wave braneworld with the bulk real scalar field, introduced in \cite{PM_2020}. It was shown earlier, in \cite{PM_2020}, that the braneworld model realizes pure gravitational trapping mechanism for particles and light and for massless scalar fields on the brane. Despite that, in this article we have explicitly shown that there is no normalizable fermion field zero mode trapped on the brane.



\begin{thebibliography}{}

\bibitem{AHDD}
N. Arkani-Hamed, S. Dimopoulos and G. Dvali,
                  Phys. Lett. B 429 (1998) 263,
                  arXiv: hep-ph/9803315.

\bibitem{AAHDD}
I. Antoniadis, N. Arkani-Hamed, S. Dimopoulos and G. Dvali,
                  Phys. Lett. B 436 (1998) 257,
                  arXiv: hep-ph/9804398.


\bibitem{G1}
M. Gogberashvili,
                  Int. J. Mod. Phys. D 11 (2002) 1635,
                  arXiv: hep-ph/9812296.

\bibitem{G2}
M. Gogberashvili,
                  Mod. Phys. Lett. A 14 (1999) 2025,
                  arXiv: hep-ph/9904383.


\bibitem{RS1}
L. Randall and R. Sundrum,
                  Phys. Rev. Lett. 83 (1999) 3370,
                  arXiv: hep-ph/9905221.


\bibitem{RS2}
L. Randall and R. Sundrum,
                  Phys. Rev. Lett. 83 (1999) 4690,
                  arXiv: hep-th/9906064.

\bibitem{Rub}
V.A. Rubakov,
                  Phys. Usp. 44 (2001) 871 (Usp. Fiz. Nauk 171 (2001) 913).


\bibitem{Lan}
D. Langlois,
                  Prog. Theor. Phys. Suppl. 148 (2003) 181,
                  arXiv: hep-th/0209261.

\bibitem{Man}
P.D. Mannheim,
                  Brane-localized Gravity. World Scientific, Singapore 2005.



\bibitem{Maa-Koy}
R. Maartens and K. Koyama,
                  Living Rev. Rel. 13 (2010) 5,
                  arXiv: 1004.3962 [hep-th].



\bibitem{Gut-Str}
M. Gutperle and A. Strominger,
                  JHEP 0204 (2002) 018,
                  arXiv: hep-th/0202210.

\bibitem{Kru-Mye-Pee}
M. Kruczenski, R.C. Myers and A.W. Peet,
                  JHEP 0205 (2002) 039,
                  arXiv: hep-th/0204144.

\bibitem{Iva-Sin}
V.D. Ivashchuk and D. Singleton,
                  JHEP 0410 (2004) 061,
                  arXiv: hep-th/0407224.


\bibitem{Bur-Que-Rab-Tas-Zav}
C.P. Burgess, F. Quevedo, R. Rabadan, G. Tasinato and I. Zavala,
                  JCAP 0402 (2004) 008,
                  arXiv: hep-th/0310122.

\bibitem{Gog-Sin1}
M. Gogberashvili and D. Singleton,
                  Mod. Phys. Lett. A 25 (2010) 2131,
                  arXiv: 0904.2828 [hep-th].

\bibitem{Gog-Her-Mal}
M. Gogberashvili, A. Herrera-Aguilar and D. Malagon-Morejon,
                  Class. Quant. Grav. 29 (2012) 025007,
                  arXiv: 1012.4534 [hep-th].


\bibitem{Gog-Her-Mal-Mor-Nuc}
M. Gogberashvili, A. Herrera-Aguilar, D. Malagon-Morejon, R.R. Mora-Luna and U. Nucamendi,
                  Phys. Rev. D 87 (2013) 084059,
                  arXiv: 1201.4569 [hep-th].

\bibitem{Gog-Her-Mal-Mor}
M. Gogberashvili, A. Herrera-Aguilar, D. Malagon-Morejon and R.R. Mora-Luna,
                  Phys. Lett. B 725 (2013) 208,
                  arXiv: 1202.1608 [hep-th].


\bibitem{Sou-Sil-Alm}
L.J.S. Sousa, J.E.G. Silva and C.A.S. Almeida,
                  arXiv: 1209.2727 [hep-th].

\bibitem{PM}
P. Midodashvili,
                  Int.J.Theor.Phys. 53 (2014) 1174,
                  arXiv: 1211.0206 [hep-th].

\bibitem{GMM1}
M. Gogberashvili, P. Midodashvili and L. Midodashvili,
                  Phys. Lett. B 702 (2011) 276,
                  arXiv: 1105.1701 [hep-th].
\bibitem{GMM2}
M. Gogberashvili, P. Midodashvili and L. Midodashvili,
                  Phys. Lett. B 707 (2012) 169,
                  arXiv: 1105.1866 [hep-th].
\bibitem{GMM3}
M. Gogberashvili, P. Midodashvili and L. Midodashvili,
                  Int. J. Mod. Phys. D 21 (2012) 1250081,
                  arXiv: 1209.3815 [hep-th].

\bibitem{G3}
M. Gogberashvili,
                  JHEP 09 (2012) 056,
                  arXiv: 1204.2448 [hep-th].


\bibitem{Gog-Sakh-Tukh}
M. Gogberashvili, O. Sakhelashvili and G. Tukhashvili,
                  Mod. Phys. Lett. A 28 (2013) 1350092,
                  arXiv: 1304.6079 [hep-th].


\bibitem{GM_RealScField1}
M. Gogberashvili and P. Midodashvili,
                  Adv.High Energy Phys. 2013 (2013) 873686,
                  arXiv: 1310.1911 [hep-th].

\bibitem{GM_RealScField2}
M. Gogberashvili and P. Midodashvili,
                  EPL 104 (2013) 50002,
                  arXiv: 1312.6241 [hep-th].

\bibitem{GM_RealScField3}
M. Gogberashvili and P. Midodashvili,
                  Int.J.Mod.Phys. A 29 (2014) 1450141,
                  arXiv: 1409.4408 [physics.gen-ph].

\bibitem{SCA}
L.J.S. Sousa, W.T. Cruz and C.A.S. Almeida,
                  Phys.Rev. D 89 (2014) 064006,
                  arXiv: 1311.5848 [hep-th].

\bibitem{PM_2020}
P. Midodashvili and L. Midodashvili,
                  Braz.J.Phys. 50 (2020) 6, 750.

\end{thebibliography}
\end{document}